\documentclass[shortnote]{jpsj2} %% for short notes
\title{Free energy of disordered urn models in the canonical ensemble}

\author{Jun \textsc{Ohkubo}\thanks{E-mail address: ohkubo@issp.u-tokyo.ac.jp}}
\inst{
Institute for Solid State Physics, University of Tokyo, 
Kashiwanoha 5-1-5, Kashiwa, Chiba 277-8581, Japan
}

\abst{
}

\kword{disordered urn model, free energy, replica analysis, random field Ising model}

\begin{document}
\maketitle

Urn models have been studied a lot because they are analytically tractable and contain rich physical phenomena,
e.g., slow relaxation or condensation phenomena
\cite{Ritort1995,Bialas1997,Drouffe1998,Godreche2001,Leuzzi2002}.
It has been revealed that the urn models
are related to zero range processes and asymmetric simple exclusion processes \cite{Evans2005}
which are stochastic processes for studying nonequilibrium statistical physics.
Furthermore, the urn models have been used in research fields of complex networks \cite{Albert2002,Ohkubo2005c}.

Recently, the analytical treatment for disordered versions of the urn models have been developed
\cite{Leuzzi2002,Ohkubo2005c,Ohkubo2006}.
Leuzzi and Ritort have analyzed a disordered urn model (disordered backgammon model)
in the scheme of the grand canonical ensemble.
On the other hand, in the scheme of the canonical ensemble,
the replica method has been used in order to calculate the occupation distribution 
of the preferential urn model\cite{Ohkubo2006}.
The replica method is a powerful tool for random systems,
but still has an ambiguity for the mathematical validation.
Furthermore, the analysis in ref.~9 has been based on replica symmetric solutions,
so that it was not clear that the replica symmetric solution is adequate for the disordered urn model,
although the solutions are in good agreement with numerical experiments.

In this short note,
we calculate the free energy of the disordered urn model using the law of large numbers.
It is revealed that the saddle point equation obtained by the usage of the law of large numbers
is the same as that obtained by the replica method in ref.~9.
Hence, we conclude that the replica symmetric solution is adequate for the disordered urn model.
Furthermore, we point out the mathematical similarity
of free energies between the urn models and the Random Field Ising Model (RFIM);
this similarity gives an evidence that the replica symmetric solution of the urn models is exact. 

\begin{figure}
\begin{center}
  \includegraphics[width=7cm,keepaspectratio,clip]{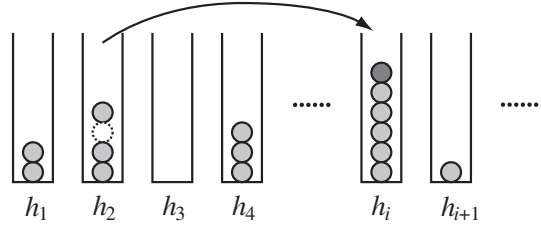} 
\caption{
Illustration of the urn model.
Each urn has a disorder parameter $h_i$.
}
\label{fig_1}
\end{center}
\end{figure}

Firstly, we give a brief explanation of a general urn model (Fig.~\ref{fig_1}).
An Urn model consists of many urns and balls.
We consider a system of $M$ balls distributed among $N$ urns;
the number of balls contained in urn $i$ is denoted by $n_i$, and hence $\sum_{i=1}^N n_i = M$.
The density of the system is defined by $\rho = M / N$.
There are two types of urn models: the Ehrenfest class and the Monkey class \cite{Godreche2001}.
In the Ehrenfest class, balls within an urn are distinguishable.
In contrast, the Monkey class has indistinguishable balls.
These two types of urn models are treated in the similar analytical method,
so that we discuss here only the case of the Ehrenfest class.
The dynamics of the Ehrenfest urn model is briefly summarized as follows:
\begin{enumerate}
\item Choose a ball at random.
\item Select an urn with a transition probability $u_i$.
\item Transfer the chosen ball to the selected urn $i$.
\item Repeat the above procedures until the system reaches an equilibrium state.
\end{enumerate}
Selecting an arbitrary transition probability $u_i$,
one can construct various urn models suitable for own physical problems.

The total energy of the whole system is defined as
\begin{align}
E(n_1, \dots, n_N) = \sum_{i=1}^N E(h_i,n_i),
\end{align}
where $E(h_i,n_i)$ is an energy of urn $i$, and $h_i$ the disorder parameter for urn $i$.
It is assumed that the disorder parameter $h_i$ is assigned to each urn by using a probability density $\phi(h)$.
The partition function of the urn model is written as\cite{Godreche2001}
\begin{align}
Z = \sum_{n_1 = 0}^\infty \cdots \sum_{n_N = 0}^\infty p(h_1,n_1) \cdots p(h_N,n_N) 
\delta(\sum_{i=1}^N n_i, M),
\label{eq_urn_partition_function}
\end{align}
where
\begin{align}
p(h_i,n_i) = \frac{1}{n_i!}e^{-\beta E(h_i,n_i)}
\end{align}
is the Boltzmann weight attached to urn $i$, and $\beta$ an inverse temperature.
The Kronecker delta $\delta(\sum_{i=1}^N n_i, M)$ stems from the constraint of the total number of balls.
Using the Boltzmann weight, the transition probability is defined as\cite{Evans2005}
\begin{align}
u_i = \frac{p(h_i,n_i-1)}{p(h_i,n_i)}.
\end{align}

The aim of the present paper is to calculate the free energy in the thermodynamic limit
by using the canonical partition function $Z$ of eq.~\eqref{eq_urn_partition_function}.
In order to calculate the free energy,
we rewrite the partition function $Z$ as
\begin{align}
Z 
&= \oint \frac{d z}{2\pi} \frac{1}{z} \exp\left[ - N \left\{ \rho \ln z - \frac{1}{N} \sum_{i=1}^N \ln H(h_i,z) 
\right\} \right],
\end{align}
where we use the integral representation of the Kronecker delta
\begin{align}
2i \pi \delta(m,n) = \oint d z \, z^{m-n-1} 
\end{align}
and define
\begin{align}
H(h,z) =  \sum_{n=0}^\infty p(h,n) z^n.
\end{align}
Hence, the free energy of the disordered urn model in the configurational average is written by
\begin{align}
&\frac{1}{N} \langle F \rangle_{\{ \mib{h} \}} 
= \frac{1}{N} (-\frac{1}{\beta}) \langle \ln Z \rangle_{\{ \mib{h} \}} \notag \\
&= - \frac{1}{\beta N} \left\langle \ln
\oint \frac{d z}{2\pi} \frac{1}{z} \exp\left[ - N \left\{ \rho \ln z 
- \frac{1}{N} \sum_{i=1}^N \ln H(h_i,z) 
\right\} \right]
\right\rangle_{\{ \mib{h}\}} 
\label{eq_free_energy_urn_model_pre},
\end{align}
where the configurational average is defined as
\begin{align}
&\left\langle A(h_1,\cdots,h_N) \right\rangle_{\{ \mib{h} \}} \notag \\
&\quad = \int d h_1 \phi(h_1) \cdots  \int d h_N \phi(h_N) A(h_1, \cdots, h_N).
\end{align}

The law of large numbers is available 
in order to calculate the configurational average of eq.~\eqref{eq_free_energy_urn_model_pre}.
Using the law of large numbers, $\frac{1}{N} \sum_{i=1}^N x_i = \left\langle x \right\rangle \quad (N \to \infty)$,
the term $(\sum_{i=1}^N \ln H(h_i,z) )/N$ in eq.~\eqref{eq_free_energy_urn_model_pre}
is replaced by $\langle \ln H(h,z) \rangle_h$.
Hence, it becomes easy to take the configurational average.
By using the saddle-point method, we get the free energy
\begin{align}
\frac{1}{N} \langle F \rangle_{\{{h}\}}
= - \frac{\rho}{\beta} \ln z_\mathrm{s} + \frac{1}{\beta} \left\langle \ln H(h,z_\mathrm{s}) \right\rangle_h
\label{eq_free_energy_urn_model}
\end{align}
and the saddle-point equation
\begin{align}
\rho = \left\langle \frac{z_\mathrm{s}}{H(h,z_\mathrm{s})} 
\frac{d}{d z_\mathrm{s}} H(h,z_\mathrm{s}) \right\rangle_h.
\label{eq_saddle_point_urn_model}
\end{align}
We note that the final result is consistent with the analytical results obtained
by the replica method \cite{Ohkubo2006}.
Hence, we conclude that the replica symmetric solution in ref.~9 is valid
for the disordered urn model.

Finally, we will point out the mathematical similarity between the urn model and the RFIM.
The RFIM consists of $N$ Ising spins interacting through an infinite ranged exchange interaction.
The Hamiltonian is given by
\begin{align}
\mathcal{H} = - \frac{J}{N} \sum_{i < j} S_i S_j - \sum_i h_i S_i ,
\end{align}
where $J$ is the spin coupling constant, $\{S_i\}$ are spin variables,
and $\{\mib{h}\} = \{h_1, \cdots, h_N \}$ are the external magnetic fields.
The free energy of the RFIM is written as\cite{Schneider1977}
%\begin{align}
%&\frac{1}{N} \langle F_\mathrm{RFIM} \rangle_{\{\mib{h}\}} \notag \\
%&= \frac{1}{\beta N} \left( \ln \sqrt{\frac{\beta JN}{2\pi}} - \frac{1}{2} \beta J\right) \notag \\
%&\quad - \frac{1}{\beta N} \left\langle
%\ln \int_{-\infty}^\infty dm \,\,
%\exp\left[ 
%-N \left\{ \frac{1}{2} \beta J m^2 \right. \right. \right. \notag \\
%&\quad \left. \left. \left. -\frac{1}{N} \sum_{i=1}^N \ln 2 \cosh \beta(h_i + Jm) \right\}
%\right]   
%\right\rangle_{\{\mib{h}\}}, \nonumber \\
%\label{eq_RFIM_free_energy_pre}
%\end{align}
\begin{align}
&\frac{1}{N} \langle F_\mathrm{RFIM} \rangle_{\{\mib{h}\}} \notag \\
&= \frac{1}{\beta N} \left( \ln \sqrt{\frac{\beta JN}{2\pi}} - \frac{1}{2} \beta J\right) \notag \\
&\quad - \frac{1}{\beta N} \left\langle
\ln \int_{-\infty}^\infty dm \,\,
\exp\left[ 
-N \left\{ \frac{1}{2} \beta J m^2
-\frac{1}{N} \sum_{i=1}^N \ln 2 \cosh \beta(h_i + Jm) \right\}
\right]   
\right\rangle_{\{\mib{h}\}}, \nonumber \\
\label{eq_RFIM_free_energy_pre}
\end{align}
where $\beta$ is the inverse temperature.
For the RFIM, it has been revealed that the replica symmetric solution gives an exact result\cite{Schneider1977}.
We can easily see the similarity of the free energy of the RFIM in eq.~\eqref{eq_RFIM_free_energy_pre} and 
that of the urn model in eq.~\eqref{eq_free_energy_urn_model_pre}:
the configurational average of a logarithm of an integral is taken,
and the term depending on the disorder parameters is contained in the integral 
as the argument of the exponential function.
We here remark a difference between the urn models and the RFIM.
The variable of integration in eq.~\eqref{eq_RFIM_free_energy_pre}
is the magnetization, $m = \left(\sum_{i = 1}^N S_i \right) / N$,
and that in the disordered urn model is related to the macroscopic constraint $M = \sum_{i=1}^N n_i$ 
or $\rho = \left(\sum_{i = 1}^N n_i \right) / N$.
The macroscopic observable $M$ (or $\rho$) in the disordered urn models
is a constraint, therefore the observable is known in advance.
In contrast, the magnetization $m$ in the RFIM 
is unknown in advance and constructed by microscopic variables $S_i$ with time.
Although there are such differences between the urn models and the RFIM,
the macroscopic observable constructed by microscopic variables ($S_i$ or $n_i$)
plays a key role in the both analysis.

In summary, we discussed a disordered version of a general urn model.
The free energy and the saddle point equation were obtained by the usage of the law of large numbers,
and it was clarified that the saddle point equation is the same as that of the replica symmetric solution.
In addition, we pointed out the similarity between the urn models and the RFIM.
Although we have not given the rigorous mathematical proof,
we expect from the similarity that the replica symmetric solution describes exact results in the thermodynamic limit.
We hope that the analogy with the RFIM will lead to better understanding of the disordered urn models.

%This work was supported in part by grant-in-aid for scientific research (No.~18$\cdot$5140)
%from the Ministry of Education, Culture, Sports, Science and Technology, Japan.

%\newpage %Just because of unusual number of tables stacked at end
%\bibliography{urn_PRP}% Produces the bibliography via BibTeX.

\end{document}